\documentclass[pre,twocolumn,superscriptaddress,floatfix,showpacs,showkeys]{revtex4}

\usepackage[latin2]{inputenc}
\usepackage{amsmath}
\usepackage{amssymb}
\usepackage{epsfig}
\usepackage{color}

\begin{document}

\title{Extent of force indeterminacy in packings of frictional rigid disks}

\date{\today}

\author{M.\ Reza Shaebani}
\affiliation{Department of Theoretical Physics, Budapest Univ.\ of
 Techn.\ and Econ., H-1111 Budapest, Hungary}
 \affiliation{Institute for Advanced Studies in Basic Sciences, Zanjan 45195-1159, Iran}
 \affiliation{Department of Theoretical Physics, University of Duisburg-Essen, 47048 Duisburg, Germany}
\author{Tam\'as Unger}
\affiliation{Department of Theoretical Physics, Budapest Univ.\ of
 Techn.\ and Econ., H-1111 Budapest, Hungary}
\affiliation{Solid State Research Group of the HAS, Budapest Univ.\ of
 Techn.\ and Econ.}
\author{J\'anos Kert\'esz}
\affiliation{Department of Theoretical Physics, Budapest Univ.\ of
 Techn.\ and Econ., H-1111 Budapest, Hungary}
\affiliation{Solid State Research Group of the HAS, Budapest Univ.\ of
 Techn.\ and Econ.}

\begin{abstract}
Static packings of frictional rigid particles are investigated by
means of discrete element simulations. We explore the ensemble of
allowed force realizations in the space of contact forces for a
given packing structure. We estimate the extent of force
indeterminacy with different methods. The indeterminacy exhibits a
nonmonotonic dependence on the interparticle friction coefficient.
We verify directly that larger force-indeterminacy is accompanied
by a more robust behavior against local perturbations. We also
investigate the local indeterminacy of individual contact forces.
The probability distribution of local indeterminacy changes its
shape depending on friction. We find that local indeterminacy
tends to be larger on force chains for intermediate friction. This
correlation disappears in the large friction limit.
\end{abstract}



\maketitle In packings of relatively rigid particles, elastic
deformations of the grains are typically several orders of
magnitude smaller than the grain size. Since this separation of
length scales occurs it is a natural idea to investigate the limit
case of infinite stiffness of the grains.

It is known that jammed packings of perfectly rigid particles with
finite friction coefficient are ``\emph{hyperstatic}''
\cite{JNRoux00,Silbert02}. The number of equations of mechanical
balance is smaller than the number of unknowns (components of the
interparticle forces). This makes the problem undetermined in the
sense that there are many solutions that satisfy the equilibrium
equations. Even taking constraint conditions, like Coulomb's limit
of friction and unilaterality of the contacts, into account does
not help to eliminate the indeterminacy of the contact forces;
Thus for a given packing geometry the solutions define an ensemble
of admissible force networks $\mathcal{S}$
\cite{Snoijer04,Snoijer04b}. $\mathcal{S}$ is a convex set
\cite{Unger05} in the force space $\mathcal{F}$, spanned by the
components of contact forces, and its boundaries are delimited by
constraint conditions.

The ensemble has received considerable attention since many
macroscopic properties of granular packings can be derived from
ensemble averaging over all allowed force states supposing a
uniform measure on $\mathcal{S}$
\cite{Ostojic05,Snoeijer06,Ostojic06,vanEerd07,Ostojic07,Ellenbroek07}.
Furthermore, with this technique one can disentangle the effect of
forces and texture of the packing. Mathematically, the problem of
finding the solutions of a set of undetermined equations and
constraints is of rather broad interest, e.g. in metabolic
networks \cite{Segre02,Almaas04}.

The extent of force indeterminacy in 2D random packings of
perfectly rigid disks was investigated theoretically and
numerically in \cite{McNamara04}. The indeterminacy of each
component of the contact forces was obtained, suggesting that
highly undetermined contacts are located on main force chains.
Force indeterminacy in such packings was also measured in
\cite{Unger05} where it turned out that the indeterminacy depends
nonmonotonically on the interparticle friction coefficient due to
the competition between two coexisting effects, the opening of the
Coulomb cone angle and the lowering of connectivity. In
Ref.\cite{Shaebani07} similar nonmonotonic friction dependence is
obtained for mechanical response of the granular packings to local
perturbations.

In this paper we examine whether the nonmonotonic friction
dependence of force indeterminacy remains valid also when other
methods are used to quantify the ``size'' of the solution set
$\mathcal{S}$.  We measure numerically the extent of
force-indeterminacy and the mechanical response to local
perturbations in the same packings and examine the relation
between them.  The local force indeterminacy is also studied in
this work. First, we investigate its probability distribution,
then, we compare its spatial pattern to that of the force chains
in the packing.

{\em Sampling Procedure -- } The systems we investigate are 2D
random packings of $400$ perfectly rigid disks. Periodic boundary
conditions are applied in both directions, disk radii are
uniformly distributed between $0.5$ and $1$, gravity is set to
zero and the unit of the length is set to the maximum grain
radius. Our numerical simulations consist of two steps which are
performed with the help of contact dynamics algorithm
\cite{Moreau94,Jean99,Brendel04}. First we construct static
configurations of particles. The initial dilute systems are
compressed by imposing a homogeneous confining pressure $P_0$ to
get the final static packings. The full description of our method
of constructing the homogeneous packings can be found in
\cite{Shaebani08jcp}.

\begin{figure}
\epsfig{figure=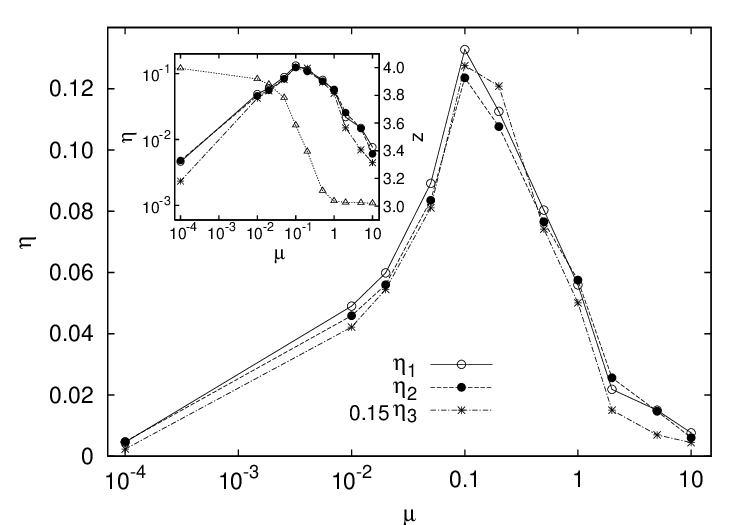,width=0.8\linewidth}
\caption{Force indeterminacy $\eta$, quantified with different
methods that are explained in the text, in terms of the friction
coefficient $\mu$. The inset displays the same plot in log-log 
scale. The average coordination number $z$ as a function of 
$\mu$ is also shown (triangles).} \label{Fig-Indeterminacy}
\end{figure}

Then we explore the force ensemble: we collect force networks that
provide static solutions for the given contact geometry and
boundary conditions. We use a random walk method in the force
space \cite{Unger03b,Unger05} starting with the original force
network. We perturb the original force state and jump to a new
force state in the force space $\mathcal{F}$. The technique is to
add random values that are chosen uniformly from the interval
$[-\langle F_n \rangle,\langle F_n \rangle]$ to all components of
the contact forces. $\langle F_n \rangle$ is the mean normal force
calculated over the current values of contact forces. The
perturbed force network is given as the input for the
Gauss-Seidel-like iterative solver of the contact dynamics method
which lets the forces relax into a new consistent state. The jump
is accepted if the new state is an equilibrium state, otherwise it
is rejected. The perturbation and relaxation are repeated many
times, always starting from the last equilibrium force network. In
this way we collect $1000$ admissible force networks for a given
static packing. In order to study systematically the influence of
the interparticle friction coefficient on the extent of the force
indeterminacy, the constructing and sampling procedures are
repeated for various values of the friction coefficient.

{\em Numerical Results -- } We first quantify the extent of the
force indeterminacy $\eta$ for a given packing geometry based on
the sampled force networks. We compare here three different
methods, since there is no a priori preferred way to measure $\eta$. 
Let us denote the center of the samples in the force
space $\mathcal{F}$ by $\{\vec G_c\}$ which is a force network
with contact force vectors $\vec G_c$ given by
\begin{equation}
   \vec G_c = \langle \vec F_c {\rangle}_\text{states}, \hspace{1cm}
   c=1,...,N_c
\label{Center-F.N.}
\end{equation}
where the average $\langle \cdot \cdot \cdot
{\rangle}_\text{states}$ is taken over all realizations of the
force states and $N_c$ is the number of contacts. One possibility
to quantify the force indeterminacy is to measure the force
fluctuations $\delta F_c$ around the mean force vector $\vec G_c$
at each contact $c$ \cite{Unger05}:
\begin{equation}
   \delta F_c =  \left\langle (\vec F_c - \vec G_c)^2
   \right\rangle^{^\text{1/2}}_{_\text{states}}.
\label{force-fluctuation}
\end{equation}
The force indeterminacy $\eta_1$ of the whole packing is given by
the relative fluctuation:
\begin{equation}
   \eta_{_1} = \frac{\langle \delta F_c {\rangle}_\text{cont.}}
   {\langle |\vec G_c| {\rangle}_\text{cont.}},
\label{Eta1}
\end{equation}
where $\langle \cdot \cdot \cdot {\rangle}_\text{cont.}$ denotes
the average over all contacts.

\begin{figure}
\epsfig{figure=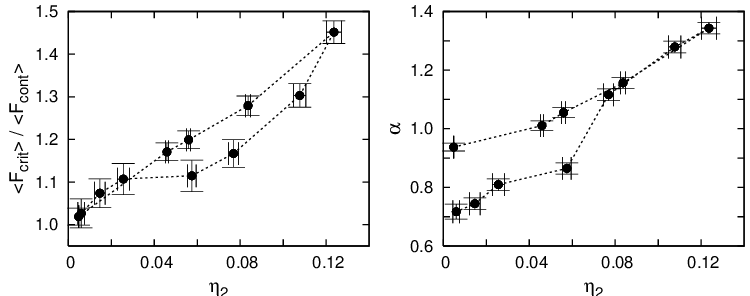,width=0.99\linewidth}
\caption{The average critical force $\langle F_\text{crit}
\rangle$ scaled by the average normal contact  force $\langle
F_\text{cont} \rangle$ (left) and the decay exponent $\alpha$
(right) as functions of the force indeterminacy $\eta_{_2}$ for
packings that are constructed with different friction
coefficients.} \label{Fig-FcritDeltaEta}
\end{figure}

The extent of the indeterminacy could be also estimated by the
Euclidean distance between randomly chosen pairs of force states
\cite{McNamara04,Snoeijer06}. The probability distribution of the
distances becomes sharply peaked if $\mathcal{S}$ is a high
dimensional object. The global indeterminacy according to this
method is defined via
\begin{equation}
   \eta_{_2} = \left(\frac{\langle (\{ \vec F_c {\}} -
   \{ \vec F_c \}^\prime)^2 {\rangle}_{_\text{pairs}}}
   {\{ \vec G_c \}^2}\right)^{^\text{1/2}}.
\label{Eta2}
\end{equation}
$\{ \vec F_c \}$ and $\{ \vec F_c \}^\prime$ are two different
force states and $\langle \cdot \cdot \cdot
{\rangle}_\text{pairs}$ means the average over all pairs of force
states. The square of a force state $\{ \vec F_c \}^2$ is given by
$\displaystyle\sum_c \vec F^2_c$.

As an alternative method \cite{McNamara04}, the extremal points of
$\mathcal{S}$ along each axis of the force space $\mathcal{F}$
provide the following measure of the indeterminacy:
\begin{equation}
   \eta_{_\text{3}} = \frac{\langle F^\text{max}-F^\text{min}
   {\rangle}_\text{comp.}}
   {\langle \frac{F^\text{max}+F^\text{min}}{2}
   {\rangle}_\text{comp.}}. \label{Eta3}
\end{equation}
Here, $F^\text{max}$ and $F^\text{min}$ are the maximum and
minimum values of a contact force component (either normal or
tangential). The average $\langle \cdot \cdot \cdot
{\rangle}_\text{comp.}$ is taken over all $2 \times \!N_c$
components of contact forces. We note that these are 
mathematically three different measures that give different 
results depending on the circumstances. E.g.\ the first two 
methods [Eqs.~(\ref{Eta1}) and (\ref{Eta2})] depend on the probability
measure that is realized by the sampling. This is not the case for
$\eta_3$ which has a pure geometrical definition (provided the
sampling explores the solution set). The question whether the
sampling is uniform or not has no effect on the value of $\eta_3$. 
Our numerical tests (not shown) reveal that the values of $\eta_i$ 
are well reproducible and do not strongly depend on the total 
number of grains.  

\begin{figure}
\epsfig{figure=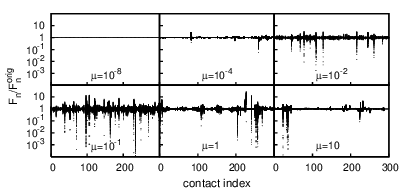,,width=0.99\linewidth}
\caption{The possible values of the normal contact forces $F_n$
(denoted by dots and mostly merged to intervals), scaled by the
normal component of the original contact force $F^\text{orig}_n$,
are shown for each contact of the packing. Each plot corresponds
to the packing with the given friction coefficient. The fluctuations 
are small for small and large $\mu$, and become larger at 
intermediate $\mu$.}
\label{Fig-FnSet}
\end{figure}

In Fig.~\ref{Fig-Indeterminacy} we compare the values of $\eta$
obtained by the three methods which, up to a constant factor,
provide basically the same behavior in the whole range of friction
($\eta_1 \approx \eta_2 \approx 0.15 \eta_3$). The nonmonotonic
friction dependence, reported in \cite{Unger05}, is reproduced
here independently of the quantifying method. We note that the 
average coordination number $z$ of the packing depends strongly 
on the friction coefficient (inset of Fig.~\ref{Fig-Indeterminacy}); 
consequently, the dimension of $\mathcal{S}$ is varied with $\mu$.

Next we investigate the effect of $\eta$ on the mechanical
response of granular packings. In
Refs.~\cite{Shaebani07,Shaebani08pre} local perturbations were
used to break the equilibrium structure of the homogeneous
packings and induce motion of the grains. It turned out that the
displacements of the particles due to local perturbations decay as
a power law of the distance from the perturbation point. The
numerical experiment was repeated for several packings constructed
with different $\mu$ revealing that the decay exponent $\alpha$
and the critical force $F_\text{crit}$, i.e. the force needed to
break the mechanical equilibrium, exhibit a nonmonotonic
dependence on the friction with extrema at $\mu = 0.1$ similarly
to the behavior of $\eta$. This similarity suggests the picture
that a packing with larger force-indeterminacy becomes more stable
against perturbations. Here, we test directly whether such a
relation exists: Together with the force indeterminacy we
determine also the response quantities $F_\text{crit}$ and
$\alpha$ for the same packing configurations. Since the different
methods we used to quantify $\eta$ are basically equivalent, we
plot the response quantities in terms of $\eta_{_2}$ in
Fig.~\ref{Fig-FcritDeltaEta}. The same series of packings are
plotted here as in Fig.~\ref{Fig-Indeterminacy}. The lines are
connecting the data points in the order of increasing friction.
Both $F_\text{crit}$ and $\alpha$ are strongly related to the
extent of force indeterminacy, although they are not a unique
function of $\eta$. Still, very different packings (with different
density, connectivity and frictional properties) exhibit similar
response properties if their $\eta$ values are close to each
other.

\begin{figure}
\epsfig{figure=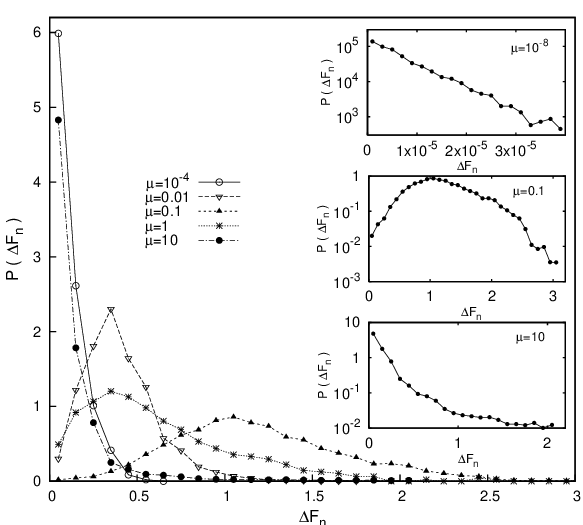,width=0.8\linewidth}
\caption{The probability distribution of the interval of possible
normal contact forces for several friction coefficients $\mu$. The
insets display semilogarithmic probability distributions for three
different $\mu$.} \label{Fig-Probability}
\end{figure}

The quantity $\eta$, as a global parameter, provides 
a good estimate of the size of $\mathcal{S}$. In order to learn more 
details on the structure of the solution set we study the 
indeterminacy also locally at the level of individual contacts. In 
Fig.~\ref{Fig-FnSet} we show the values of normal contact forces 
$F_n$ for every contact obtained by the sampled realizations of 
force states. We note that the data points are highly correlated, 
the contact forces can not vary independently of each other, but 
here we investigate only how the solution set $\mathcal{S}$ is 
elongated along each ``$F_n$" axis. At each contact the possible 
values of $F_n$ form an interval $\Delta F_n$ because these values 
are the projection of the convex solution set $\mathcal{S}$ onto 
the $F_n$-axis. The length of the interval $\Delta F_n$ serves as 
a measure of the local indeterminacy at a given contact and can be 
estimated with the help of extrema of $F_n$ that were provided by 
the sampling procedure.

\begin{figure}[b]
\epsfig{figure=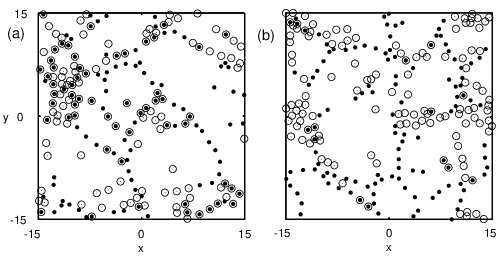,width=0.99\linewidth}
\caption{The position of the contacts with original normal force
$F_n$ larger than $2 \langle F_n \rangle$ ($\bullet$) and the
position of the contacts with force indeterminacy $\Delta F_n$
larger than $\langle \Delta F_n \rangle$ ($\odot$) in the
packing with $\mu = 0.1$ (a) and $\mu = 10$ (b).}
\label{Fig-Chains}
\end{figure}

Figure~\ref{Fig-FnSet} reveals that the fluctuations 
of the lengths $\Delta F_n$ grow nonmonotonically with friction 
similarly to the behavior of $\eta$. Moreover, it can be also seen 
from Fig.~\ref{Fig-FnSet} that the length of $\Delta F_n$ varies 
widely from one contact to another, even for a given friction 
coefficient. We now aim to investigate how frequent the different 
magnitudes of the local indeterminacies are. In Fig.~\ref{Fig-Probability} 
we show the probability distribution of $\Delta F_n$ for different 
friction coefficients. $P(\Delta F_n)$ is a monotonically decreasing 
function for small and large friction limits, but becomes broader 
and displays a peak for intermediate friction coefficients. This shows 
that the value of friction coefficient has even a geometrical signature 
in the solution set. Intermediate values of $\mu$ provide a more compact 
shape of $\mathcal{S}$ in the sense that the frequency of extreme dimensions 
are suppressed compared to the case of small and large $\mu$ where we find 
more pronounced extremes and therefore a more elongated shape of the 
solution set. The tail of $P(\Delta F_n)$ follows exponential decay for 
small frictions, while it decays faster (slower) than exponential for
intermediate (large) frictions (see the insets of Fig.~\ref{Fig-Probability}).

\begin{figure}
\epsfig{figure=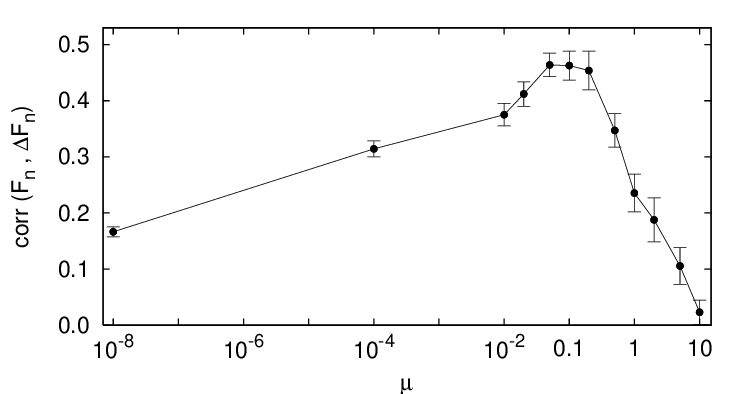,width=0.8\linewidth}
\caption{The correlation between the normal component of the
original contact force $F_n$ and the force indeterminacy
$\Delta F_n$ in terms of the friction coefficient $\mu$. The 
error bars display the maximum possible change in the 
correlation values when an arbitrary force network is chosen.}
\label{Fig-Correlation}
\end{figure}

Finally, we investigate the spatial distribution of the
indeterminacy throughout the system. The aim is to find whether
contacts that are located in a main force chain carry also larger
force indeterminacies. In Fig.~\ref{Fig-Chains} (full circles) we
plot the location of contacts that have larger $F_n$ than twice of
the average normal contact force $\langle F_n \rangle$ (according
to the original force network in the packing). We also plot the
contacts with large force-indeterminacy $\Delta F_n$ (open
circles) above the average $\langle \Delta F_n \rangle$. This way
approximately the same number of open and full circles are
plotted. It can be seen that contacts in force chains tend to have
larger force indeterminacy in case of intermediate friction
[Fig.~\ref{Fig-Chains}(a)], but for $\mu = 10$ the two patterns
become seemingly different [Fig.~\ref{Fig-Chains}(b)]. Indeed, if
we determine the correlation between $\Delta F_n$ and
$F_n$ over all contacts of the original force network and plot it 
against the friction coefficient (Fig.~\ref{Fig-Correlation}), 
it reveals that the correlation vanishes for large frictions. 
Interestingly, the correlation exhibits again a nonmonotonic 
dependence on friction, where significant correlations are 
present for the intermediate friction regime and weaker 
correlations outside. This may have a statistical origin. When 
the indeterminacy is large (intermediate friction) it is plausible 
to expect that the correlation between $\Delta F_n$ and $F_n$ is 
enhanced, since finding a large force is more probable in a large 
interval of positive normal forces \footnote{We thank the referee 
for his/her valuable suggestion.}.

We note that local force indeterminacies can be seen everywhere in
our packings. This is in contrast to what has been reported in
Ref.~\cite{Unger05}, where in case of large friction undetermined
contacts formed localized clusters while contact forces in the
rest of the packing became uniquely determined. Such localization
is absent in the present study. We think that this difference
originates from the different boundary conditions. In
\cite{Unger05} the forces were kept fixed at the boundary which
furthers the formation of a fully determined region. This is not
the case here, where we prescribe only the global pressure.

{\em Conclusion -- } In this paper we presented the numerical
results of the measurement of force indeterminacy in packings of
frictional hard disks. We quantified the global force
indeterminacy $\eta$ of the packing with different methods and
systematically studied the effect of interparticle friction
coefficient. $\eta$ depends nonmonotonically on friction. We
showed that the extent of force indeterminacy has an important
influence on the mechanical response properties of the material.

The indeterminacy was also studied locally at the level of individual
contacts. As a consequence, we observed a nonmonotonic influence of the
friction on the shape of the solution set. Small and large values of the
friction coefficient enhance the relative weight of extreme dimensions and
lead to a more elongated solution set in the force space.  We found
significant correlation between the spatial pattern of the
force-indeterminacy and force chains for intermediate friction, however,
the correlation disappeared in the large friction limit.

{\em Acknowledgments -- } We acknowledge support by grants
No.~OTKA T049403, No.~OTKA PD073172 and by the Bolyai Janos
Scholarship of the Hungarian Academy of Sciences.

\bibliography{Ref}

\begin{thebibliography}{21}
\expandafter\ifx\csname natexlab\endcsname\relax\def\natexlab#1{#1}\fi
\expandafter\ifx\csname bibnamefont\endcsname\relax
  \def\bibnamefont#1{#1}\fi
\expandafter\ifx\csname bibfnamefont\endcsname\relax
  \def\bibfnamefont#1{#1}\fi
\expandafter\ifx\csname citenamefont\endcsname\relax
  \def\citenamefont#1{#1}\fi
\expandafter\ifx\csname url\endcsname\relax
  \def\url#1{\texttt{#1}}\fi
\expandafter\ifx\csname urlprefix\endcsname\relax\def\urlprefix{URL }\fi
\providecommand{\bibinfo}[2]{#2}
\providecommand{\eprint}[2][]{\url{#2}}

\bibitem[{\citenamefont{Roux}(2000)}]{JNRoux00}
\bibinfo{author}{\bibfnamefont{J.~N.} \bibnamefont{Roux}},
  \bibinfo{journal}{Phys. Rev. E} \textbf{\bibinfo{volume}{61}},
  \bibinfo{pages}{6802} (\bibinfo{year}{2000}).

\bibitem[{\citenamefont{Silbert et~al.}(2002)\citenamefont{Silbert, Ertas,
  Grest, Halsey, and Levine}}]{Silbert02}
\bibinfo{author}{\bibfnamefont{L.~E.} \bibnamefont{Silbert}},
  \bibinfo{author}{\bibfnamefont{D.}~\bibnamefont{Ertas}},
  \bibinfo{author}{\bibfnamefont{G.~S.} \bibnamefont{Grest}},
  \bibinfo{author}{\bibfnamefont{T.~C.} \bibnamefont{Halsey}},
  \bibnamefont{and} \bibinfo{author}{\bibfnamefont{D.}~\bibnamefont{Levine}},
  \bibinfo{journal}{Phys. Rev. E} \textbf{\bibinfo{volume}{65}},
  \bibinfo{pages}{031304} (\bibinfo{year}{2002}).

\bibitem[{\citenamefont{Snoeijer
  et~al.}(2004{\natexlab{a}})\citenamefont{Snoeijer, Vlugt, van Hecke, and van
  Saarloos}}]{Snoijer04}
\bibinfo{author}{\bibfnamefont{J.~H.} \bibnamefont{Snoeijer}},
  \bibinfo{author}{\bibfnamefont{T.~J.~H.} \bibnamefont{Vlugt}},
  \bibinfo{author}{\bibfnamefont{M.}~\bibnamefont{van Hecke}},
  \bibnamefont{and} \bibinfo{author}{\bibfnamefont{W.}~\bibnamefont{van
  Saarloos}}, \bibinfo{journal}{Phys. Rev. Lett.}
  \textbf{\bibinfo{volume}{92}}, \bibinfo{pages}{054302}
  (\bibinfo{year}{2004}{\natexlab{a}}).

\bibitem[{\citenamefont{Snoeijer
  et~al.}(2004{\natexlab{b}})\citenamefont{Snoeijer, Vlugt, Ellenbroek, van
  Hecke, and van Leeuwen}}]{Snoijer04b}
\bibinfo{author}{\bibfnamefont{J.~H.} \bibnamefont{Snoeijer}},
  \bibinfo{author}{\bibfnamefont{T.~J.~H.} \bibnamefont{Vlugt}},
  \bibinfo{author}{\bibfnamefont{W.~G.} \bibnamefont{Ellenbroek}},
  \bibinfo{author}{\bibfnamefont{M.}~\bibnamefont{van Hecke}},
  \bibnamefont{and} \bibinfo{author}{\bibfnamefont{J.~M.~J.} \bibnamefont{van
  Leeuwen}}, \bibinfo{journal}{Phys. Rev. E} \textbf{\bibinfo{volume}{70}},
  \bibinfo{pages}{061306} (\bibinfo{year}{2004}{\natexlab{b}}).

\bibitem[{\citenamefont{Unger et~al.}(2005)\citenamefont{Unger, Kert\'esz, and
  Wolf}}]{Unger05}
\bibinfo{author}{\bibfnamefont{T.}~\bibnamefont{Unger}},
  \bibinfo{author}{\bibfnamefont{J.}~\bibnamefont{Kert\'esz}},
  \bibnamefont{and} \bibinfo{author}{\bibfnamefont{D.~E.} \bibnamefont{Wolf}},
  \bibinfo{journal}{Phys. Rev. Lett.} \textbf{\bibinfo{volume}{94}},
  \bibinfo{pages}{178001} (\bibinfo{year}{2005}).

\bibitem[{\citenamefont{Ostojic and Panja}(2005)}]{Ostojic05}
\bibinfo{author}{\bibfnamefont{S.}~\bibnamefont{Ostojic}} \bibnamefont{and}
  \bibinfo{author}{\bibfnamefont{D.}~\bibnamefont{Panja}},
  \bibinfo{journal}{Europhys. Lett.} \textbf{\bibinfo{volume}{71}},
  \bibinfo{pages}{70} (\bibinfo{year}{2005}).

\bibitem[{\citenamefont{Snoeijer et~al.}(2006)\citenamefont{Snoeijer,
  Ellenbroek, Vlugt, and van Hecke}}]{Snoeijer06}
\bibinfo{author}{\bibfnamefont{J.~H.} \bibnamefont{Snoeijer}},
  \bibinfo{author}{\bibfnamefont{W.~G.} \bibnamefont{Ellenbroek}},
  \bibinfo{author}{\bibfnamefont{T.~J.~H.} \bibnamefont{Vlugt}},
  \bibnamefont{and} \bibinfo{author}{\bibfnamefont{M.}~\bibnamefont{van
  Hecke}}, \bibinfo{journal}{Phys. Rev. Lett.} \textbf{\bibinfo{volume}{96}},
  \bibinfo{pages}{098001} (\bibinfo{year}{2006}).

\bibitem[{\citenamefont{Ostojic and Panja}(2006)}]{Ostojic06}
\bibinfo{author}{\bibfnamefont{S.}~\bibnamefont{Ostojic}} \bibnamefont{and}
  \bibinfo{author}{\bibfnamefont{D.}~\bibnamefont{Panja}},
  \bibinfo{journal}{Phys. Rev. Lett.} \textbf{\bibinfo{volume}{97}},
  \bibinfo{pages}{208001} (\bibinfo{year}{2006}).

\bibitem[{\citenamefont{van Eerd et~al.}(2007)\citenamefont{van Eerd,
  Ellenbroek, van Hecke, Snoeijer, and Vlugt}}]{vanEerd07}
\bibinfo{author}{\bibfnamefont{A.~R.~T.} \bibnamefont{van Eerd}},
  \bibinfo{author}{\bibfnamefont{W.~G.} \bibnamefont{Ellenbroek}},
  \bibinfo{author}{\bibfnamefont{M.}~\bibnamefont{van Hecke}},
  \bibinfo{author}{\bibfnamefont{J.~H.} \bibnamefont{Snoeijer}},
  \bibnamefont{and} \bibinfo{author}{\bibfnamefont{T.~J.~H.}
  \bibnamefont{Vlugt}}, \bibinfo{journal}{Phys. Rev. E}
  \textbf{\bibinfo{volume}{75}}, \bibinfo{pages}{060302(R)}
  (\bibinfo{year}{2007}).

\bibitem[{\citenamefont{Ostojic et~al.}(2007)\citenamefont{Ostojic, Vlugt, and
  Nienhuis}}]{Ostojic07}
\bibinfo{author}{\bibfnamefont{S.}~\bibnamefont{Ostojic}},
  \bibinfo{author}{\bibfnamefont{T.~J.~H.} \bibnamefont{Vlugt}},
  \bibnamefont{and} \bibinfo{author}{\bibfnamefont{B.}~\bibnamefont{Nienhuis}},
  \bibinfo{journal}{Phys. Rev. E} \textbf{\bibinfo{volume}{75}},
  \bibinfo{pages}{030301(R)} (\bibinfo{year}{2007}).

\bibitem[{\citenamefont{Ellenbroek and Snoeijer}(2007)}]{Ellenbroek07}
\bibinfo{author}{\bibfnamefont{W.~G.} \bibnamefont{Ellenbroek}}
  \bibnamefont{and} \bibinfo{author}{\bibfnamefont{J.~H.}
  \bibnamefont{Snoeijer}}, \bibinfo{journal}{J. Stat. Mech.:Theory Exp.}
  \textbf{\bibinfo{volume}{1}}, \bibinfo{pages}{P01023} (\bibinfo{year}{2007}).

\bibitem[{\citenamefont{Segre et~al.}(2002)\citenamefont{Segre, Vitkup, and
  Church}}]{Segre02}
\bibinfo{author}{\bibfnamefont{D.}~\bibnamefont{Segre}},
  \bibinfo{author}{\bibfnamefont{D.}~\bibnamefont{Vitkup}}, \bibnamefont{and}
  \bibinfo{author}{\bibfnamefont{G.~M.} \bibnamefont{Church}},
  \bibinfo{journal}{PNAS} \textbf{\bibinfo{volume}{99}}, \bibinfo{pages}{15112}
  (\bibinfo{year}{2002}).

\bibitem[{\citenamefont{Almaas et~al.}(2004)\citenamefont{Almaas, Kov\'acs,
  Vicsek, Oltvai, and Barab\'asi}}]{Almaas04}
\bibinfo{author}{\bibfnamefont{E.}~\bibnamefont{Almaas}},
  \bibinfo{author}{\bibfnamefont{B.}~\bibnamefont{Kov\'acs}},
  \bibinfo{author}{\bibfnamefont{T.}~\bibnamefont{Vicsek}},
  \bibinfo{author}{\bibfnamefont{Z.~N.} \bibnamefont{Oltvai}},
  \bibnamefont{and} \bibinfo{author}{\bibfnamefont{A.~L.}
  \bibnamefont{Barab\'asi}}, \bibinfo{journal}{Nature}
  \textbf{\bibinfo{volume}{427}}, \bibinfo{pages}{839} (\bibinfo{year}{2004}).

\bibitem[{\citenamefont{McNamara and Herrmann}(2004)}]{McNamara04}
\bibinfo{author}{\bibfnamefont{S.}~\bibnamefont{McNamara}} \bibnamefont{and}
  \bibinfo{author}{\bibfnamefont{H.}~\bibnamefont{Herrmann}},
  \bibinfo{journal}{Phys. Rev. E} \textbf{\bibinfo{volume}{70}},
  \bibinfo{pages}{061303} (\bibinfo{year}{2004}).

\bibitem[{\citenamefont{Shaebani et~al.}(2007)\citenamefont{Shaebani, Unger,
  and Kert\'esz}}]{Shaebani07}
\bibinfo{author}{\bibfnamefont{M.~R.} \bibnamefont{Shaebani}},
  \bibinfo{author}{\bibfnamefont{T.}~\bibnamefont{Unger}}, \bibnamefont{and}
  \bibinfo{author}{\bibfnamefont{J.}~\bibnamefont{Kert\'esz}},
  \bibinfo{journal}{Phys. Rev. E} \textbf{\bibinfo{volume}{76}},
  \bibinfo{pages}{030301(R)} (\bibinfo{year}{2007}).

\bibitem[{\citenamefont{Moreau}(1994)}]{Moreau94}
\bibinfo{author}{\bibfnamefont{J.~J.} \bibnamefont{Moreau}},
  \bibinfo{journal}{Eur. J. Mech. A/Solids} \textbf{\bibinfo{volume}{13}},
  \bibinfo{pages}{93} (\bibinfo{year}{1994}).

\bibitem[{\citenamefont{Jean}(1999)}]{Jean99}
\bibinfo{author}{\bibfnamefont{M.}~\bibnamefont{Jean}},
  \bibinfo{journal}{Comput. Methods Appl. Mech. Eng.}
  \textbf{\bibinfo{volume}{177}}, \bibinfo{pages}{235} (\bibinfo{year}{1999}).

\bibitem[{\citenamefont{Brendel et~al.}(2004)\citenamefont{Brendel, Unger, and
  Wolf}}]{Brendel04}
\bibinfo{author}{\bibfnamefont{L.}~\bibnamefont{Brendel}},
  \bibinfo{author}{\bibfnamefont{T.}~\bibnamefont{Unger}}, \bibnamefont{and}
  \bibinfo{author}{\bibfnamefont{D.~E.} \bibnamefont{Wolf}}, in
  \emph{\bibinfo{booktitle}{The Physics of Granular Media}}
  (\bibinfo{publisher}{Wiley-VCH}, \bibinfo{address}{Weinheim},
  \bibinfo{year}{2004}), pp. \bibinfo{pages}{325--343}.

\bibitem[{\citenamefont{Shaebani
  et~al.}(2008{\natexlab{a}})\citenamefont{Shaebani, Unger, and
  Kert\'esz}}]{Shaebani08jcp}
\bibinfo{author}{\bibfnamefont{M.~R.} \bibnamefont{Shaebani}},
  \bibinfo{author}{\bibfnamefont{T.}~\bibnamefont{Unger}}, \bibnamefont{and}
  \bibinfo{author}{\bibfnamefont{J.}~\bibnamefont{Kert\'esz}}
  (\bibinfo{year}{2008}{\natexlab{a}}), \bibinfo{note}{arXiv:0803.3566
  [physics.comp-ph]}.

\bibitem[{\citenamefont{Unger and Kert\'esz}(2003)}]{Unger03b}
\bibinfo{author}{\bibfnamefont{T.}~\bibnamefont{Unger}} \bibnamefont{and}
  \bibinfo{author}{\bibfnamefont{J.}~\bibnamefont{Kert\'esz}},
  \bibinfo{journal}{Int. J. of Mod. Phys. B} \textbf{\bibinfo{volume}{17}},
  \bibinfo{pages}{5623} (\bibinfo{year}{2003}).

\bibitem[{\citenamefont{Shaebani
  et~al.}(2008{\natexlab{b}})\citenamefont{Shaebani, Unger, and
  Kert\'esz}}]{Shaebani08pre}
\bibinfo{author}{\bibfnamefont{M.~R.} \bibnamefont{Shaebani}},
  \bibinfo{author}{\bibfnamefont{T.}~\bibnamefont{Unger}}, \bibnamefont{and}
  \bibinfo{author}{\bibfnamefont{J.}~\bibnamefont{Kert\'esz}},
  \bibinfo{journal}{Phys. Rev. E} \textbf{\bibinfo{volume}{78}},
  \bibinfo{pages}{011308} (\bibinfo{year}{2008}{\natexlab{b}}).

\end{thebibliography}

\end{document}